# CO-induced lifting of Au (001) surface reconstruction

M.S. Pierce[1], K-C Chang[1], D.C. Hennessy[1], V. Komanicky[1,2], A. Menzel[1,3], H. You[1]

[1]Materials Science Division, Argonne National Laboratory, Argonne IL 60439, USA,

[2]Institut for Experimental Physics, Slovak Academy of Science Košice, Slovakia,

[3]Paul Scherrer Institute, CH-5232 Villigen PSI, Switzerland

**Abstract**

We report CO-induced lifting of the hexagonal surface reconstruction on Au (001). Using *in-situ* surface x-ray scattering, we determined a pressure-temperature phase diagram for the reconstruction and measured the dynamical evolution of the surface structure in real time. Our observations provide evidence that, under certain conditions, even macroscopic Au surfaces, much larger than catalytic Au nanoparticles [M. Haruta*, Catal. Today* **36**, 153 (1997)], can exhibit some of the reactive properties and surface transitions observed in systems known to be catalytically active such as Pt (001).

Gold nanoparticles under 5 nm demonstrate surprisingly large catalytic activity[1]. While the catalytic activity stems mainly from the quantum size effect,[2] one may question whether a large Au crystal surface can exhibit some of the reactive properties and surface transitions observed in systems known to be catalytically active such as Pt (001). In this letter, we demonstrate how the hexagonal reconstruction of Au (001) surface is in fact more interactive with CO than previously thought. Using *in-situ* surface x-ray scattering (SXS), we show that a brief exposure of CO gas of as little as $10^{-3}$ atm pressure lifts the hex reconstruction.

The corrugated hexagonal, or simply 'hex', reconstruction of Pt (001) surface is known to lift in the presence of carbon-monoxide[3] (CO). This generated a great deal of interest in the surface science community because of its implications both for nano-scale structuring and heterogeneous catalysis. When the Pt (001) surface reconstruction lifts, it leaves a series of bulk terminated nano-islands elongated along the (110) lattice directions. Like Pt (001), the Au (001) surface also reconstructs into very similar corrugated hexagonal rows. This behavior, first observed with low-energy electron diffraction (LEED)[4], was extensively studied as a function of temperature in ultra high vacuum (UHV) conditions.[5]

Our samples were cylindrical Au single crystals, 6 mm in diameter and 4-6 mm tall, oriented and polished over the (001) crystal plane. The scattering experiments were performed at beamline 11ID-D at the Advanced Photon Source at Argonne National Laboratory. The samples were mounted on a quartz holder set on a 6-circle diffractometer[6] and enclosed in thin-walled quartz tubing, permitting control of the gas environment while allowing transmission of x-rays. Experiments were performed in standard SXS geometry[5,6]. We controlled the flow of gas through the system using calibrated mass flow controllers. Typically the Ar gas was balanced[7] with ~ 4% $H_2$ and the samples were grounded to minimize the radiation damage and charge buildup on the hex reconstruction. The total pressure including CO was always 1 atm. A radio frequency (RF) induction heater was used to control the sample temperature *in situ* during the experiments. This enabled control of the temperature over the entire range from room temperature to the Au bulk melting point. The sample temperature was monitored by the positions of the bulk diffraction peaks during the experiments using thermal expansion data of pure gold[8]. The uncertainty in our temperature measurements is much better than 1% (~3 K) near room temperature and increases to ~2% (~20 K) at the highest temperatures of our measurements.

We took care during the experiments to ensure the results were independent of x-ray exposure,[7,9] $H_2$ presence, and sample history. We tested the x-ray effects by frequently interrupting or reducing the x-ray exposure during our measurements. The results discussed below, such as a restoration of the hex phase, were unaffected by such tests, indicating no significant x-ray-induced damage or carbon deposition. Additionally, we performed low-energy electron diffraction (LEED) and x-ray photo-emission spectroscopy (XPS) experiments in an ultra-high vacuum (UHV) chamber. First, we verified the hex reconstruction with LEED and no carbon or oxygen signals with XPS. Then the samples were exposed to 1 atm of Ar or CO for 30 seconds. Subsequent LEED measurements showed that exposure to Ar had little effect on the reconstruction while CO completely lifted the reconstruction, while XPS measurements showed still no C or O signals.

Prior to the experiments the samples were cleaned in nitric acid and degreased. X-ray reflectivity measurements of the surface before annealing usually showed poor reflectivity. After annealing at 95% of the bulk melting point we found a strong surface signal. This high temperature was frequently exploited to clean and "reset" the sample to a known initial condition, which seems justified in particular because the system, thus prepared, exhibited little difference from the properties seen in UHV conditions[5] to establish a history independent starting point for our observations.

There are two possible orientations for hex domains on the (1×1) fcc surface and each of these orientations has six equivalent first-order surface scattering peaks as shown in the insets of Figure 1. We thus use the (1.21 1.21 0) peak as representative for the hex reconstruction. At an incident flux of ~ $10^{12}$ photons/sec at 11.5 keV on the 6 mm disk sample, with a full-with-half-maxima beam profile of 0.7 mm by 0.25 mm, the observed intensity at the (1.21 1.21 0.3) surface reconstruction peak was ~ $10^3$-$10^4$ photons/sec.

First we established that there was relatively little change by performing the experiment in an Ar-$H_2$ gas, compared to previous UHV studies of the Au (001) system. An Ar-$H_2$ gas environment was used because we expected relatively little or no chemical interaction with the surface.[7] We observed the reconstruction to lift at temperatures higher than 1170 K. Below this temperature, the hex reconstruction persisted all the way down to room temperature. Our (00L) and (1.21 1.21 L) scans confirmed a single hexagonal layer of Au atoms on the surface, like in the UHV studies. However, our hexagonal domains did not continuously rotate as the temperature decreased as observed in UHV. In additional experiments with pure Ar, He, and dry-air environments we found results similar to our Ar-$H_2$ experiments. This correspondence to UHV studies provides a useful baseline, allowing us to prepare our sample surface *in situ*, and permitting subsequent comparison of changes with already known behavior.

At room temperature, we introduced CO at a partial pressure of ~1%, and observed the hex peak signal disappeared immediately. The hex peak intensities before and after this are shown in Figure 1. The intensity decreased at the anti-Bragg position (001) as well. This confirms that there is no longer a smooth surface, consistent with the presence of extra atoms on the surface from lifting of the hex phase. Subsequent annealing in 100% Ar-$H_2$ restored the hex phase. Next, starting from high temperature, above 1170 K, we slowly cooled the sample in the presence of 2.5% CO. After each temperature change we waited several minutes to ensure the system stabilized. At temperatures above 1170 K there was no observable effect by CO. Between 1170 K and 500 K we observed the hex phase, albeit with decreased intensity, as can be seen between the black and blue curves in Figure 2. The reduction in scattering intensity is too large to be caused by the interference of adsorbed CO on the surface. Instead, this result can only be explained by CO partially lifting the hex phase. Below 500 K the hex reconstruction is completely lifted with 2.5% CO. Upon heating, however, the hex phase does not return until the temperature reaches ~ 670 K. This hysteresis effect is shown in Figure 2 by the descending open circles (blue) and the ascending x markers (orange).

We recorded the hex peak intensity along several isobars and isotherms to determine the phase diagram, Figure 3. The curved diagonal dashed line indicates a constant chemical potential line given by $P \propto P_0 e^{-\Delta\mu/k_B T}$ where $\Delta\mu$ is 0.35(10) eV when it is assumed to be independent of T. We note that this line also serves as a guide to the eyes indicating the phase boundary between CO induced (1×1) and hex phase. The value 0.35(10) eV then provides an estimation of the energy difference per CO molecule between the hex+$CO_{gas}$ and (1×1)+$CO_{ad}$. The boundaries shown by vertical lines between Phases II and III at 670 K and Phases III and IV at 1170 K are independent of CO partial pressure. Therefore, we may conclude that these boundaries are largely unaffected by CO interaction with the surface. It is remarkable that restoration of the hex phase upon heating is independent of CO partial pressures while lifting the hex phase depends strongly on CO partial pressures as we can see by the hysteresis shown in Figure 2 and the curved phase boundary in Figure 3.

We also measured the dynamic behavior of the surface phase transition in real time. To measure the transition rate, we centered the detector on either the (1.2, 1.2, 0.3) hex peak or the (001) anti-Bragg peak, adjusted the system temperature or CO partial pressure, and then took data as the system evolved. An example of the observed hex peak intensity as a function of time is shown in the inset of Figure 4. These data are taken by introducing 3.3% CO to lift hex phase and subsequently stopping the CO flow to restore the hex phase. We found that decay of the hex phase follows a simple exponential behavior, $I = I_0 e^{-rt}$ where $r$ is the hex to (1×1) transition rate and $t$ is time. While this form implies random nucleation, it is presumably from a large number of step edges and kinks on the crystal surface[10]. The transition rates obtained by fitting to exponentials are shown as a function of inverse temperature in Figure 4. The transition rate *decreases* with temperature, indicating the transition is not limited by an activation barrier. Rather, it depends on CO coverage, similar to the Pt (001) case[11]. Thus it is likely that the transition rate is a result of two processes: i) CO adsorption on the surface and ii) hex to (1×1) transition catalyzed by the presence of CO. In a small-coverage limit, the

CO coverage is proportional to $T^{-\frac{1}{2}}e^{E_2/k_B T}$ where $E_2$ is the CO adsorption energy.[12]

Then the transition rate is proportional to $T^{-\frac{1}{2}}e^{(E_2-A_2)/k_B T}$ where $A_2$ is the activation energy associated with the hex to (1×1) transition with CO. This explains qualitatively the dependence of the transition rate to the inverse temperature. In this case the fit value of 0.28(8) eV is the adsorption energy of CO ($E_2$) minus the activation barrier ($A_2$), providing a lower bound for the CO adsorption energy. When we combine this result with Δμ = 0.35(10) eV discussed above (see the energy level diagram inset of Figure 4), we can conclude that the activation barrier ($A_2$) and the energy difference between hex+$CO_{ad}$ and (1×1)+$CO_{ad}$ are probably as small as ~0.1 eV.

For the growth of the hex phase, the transition rate *increases* with temperature, but is independent of CO partial pressure. Therefore, the hex growth can be explained by introducing an activation barrier independent of CO adsorption. In this case, the transition rate has a simple exponential dependence, $e^{-A_1/k_B T}$, where $A_1$ is the activation energy of the (1×1) to hex transition. Then we obtain the activation energy from the fit value, $A_1$=0.54(11) eV. This value is approximately half the corresponding value for the Pt (001) system[11]. The activation barrier is justified by a similar mechanism as described in DFT calculations of van Beurden *et al* [13]. However, we observe that a (1×1) to hex transition temperature is similar, though slightly higher, to that seen in the Pt (001) system, implying the energy difference between the hex and (1×1) phases in Au (001) is larger than 0.2 eV[11]. In this case, a ratio of the temperatures would imply an energy difference of ~ 0.25 eV. It would be interesting to see similar DFT calculations applied to the Au (001) system.

Lifting of the hex phase by CO adsorption has been extensively studied in the case of Pt (001)[3,14] and explained in subsequent theoretical work using Monte Carlo[15] and Molecular Dynamics simulations[13]. However, lifting of the hex phase by CO in Au (001) remained unknown before and no parallel studies exist. In fact, the previous studies of CO and NO on Au (001)[16] reported that CO did not lift the reconstruction while NO did in a similar experimental condition at low pressures and temperatures. This result was understandable since Au is known to be inert. The filled $5d$ shell and the shielded $6s$ shell[17] makes Au to react only with significantly more active molecules, such as NO with an unpaired electron. On the other hand, Pt with the unfilled $5d$ shell tends to exhibit a considerable reactivity and a unique catalytic activity. Therefore, CO is known to adsorb directly on the Pt (001) hex and bulk surfaces, while in the case of Au, CO is observed not to persist on the surface in experiments such as Temperature Programmed Desorption (TPD) work[18].

More recently, however, it was reported that CO can lift the reconstructions on Au (110) and Au(111) at sufficiently high pressures.[19] These reports and our results provide a possibility that there might be a considerable parallelism between Au and Pt in spite of the well-established differences. Then one may question whether the origin of this parallelism might even be responsible in part for the catalytic activities of gold nanoparticles.[1]

In summary, we have shown how the hex reconstruction of Au (001) surface is lifted in CO at higher pressure than previously studied.[16] In a series isotherm and isobar measurements, we obtained a phase diagram and an estimation for CO binding energy to Au (001) surface. We suggest the interaction of the CO and hex surface must be quite subtle, involving perhaps the temporary adsorption and low coverage of CO causing the nucleation of the (1×1) phase and its subsequent growth or the inhibition of the (1×1) to hex transition. We find considerable parallel behaviors between Au (001) and Pt (001), and our work offers another benchmark for theoretical studies of the reconstructions on 5$d$ metals, such as the molecular dynamics modeling by van Beurden et al[13].

We would like to acknowledge the kind help and assistance from Dr. Klaus Attenkoffer at the beamline during our experiments. We also benefited from many useful discussions with N. Marković, Z. Nagy, and V. Stamenković. This work was supported by the U.S. Department of Energy, Office of Basic Energy Sciences, Materials Science Division, and the use of the Advanced Photon Source was supported by the U.S. Department of Energy, Office of Basic Energy Sciences, under DOE Contract No. DE-AC02-06CH11357.

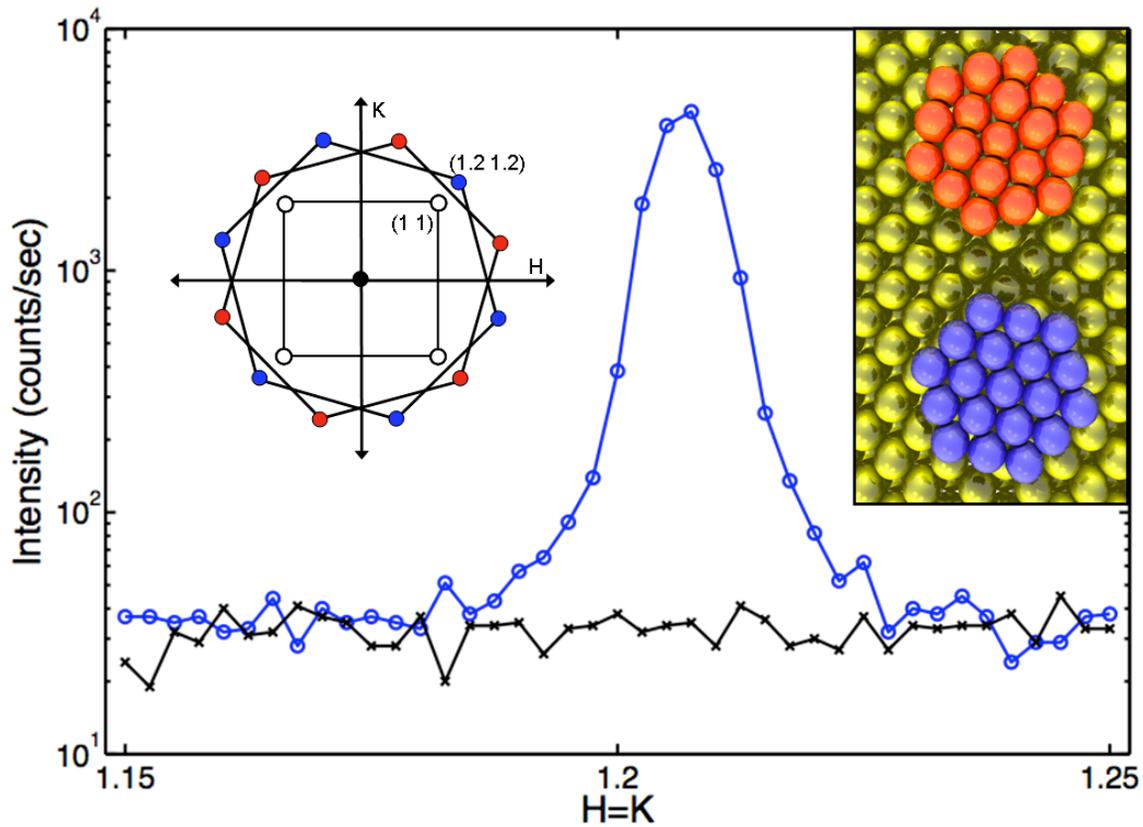

Figure 1. H=K scans across the hex peak at (1.21 1.21 0.3) of the sample annealed in Ar gas (blue circles) and after the introduction of 1.7% CO (black x marks). Inset Left: Reciprocal space diagram for L = 0. There are 12 equivalent {1.21 1.21} hex peaks. Inset Right: Two domains of the hex phase over the fcc (001) bulk.

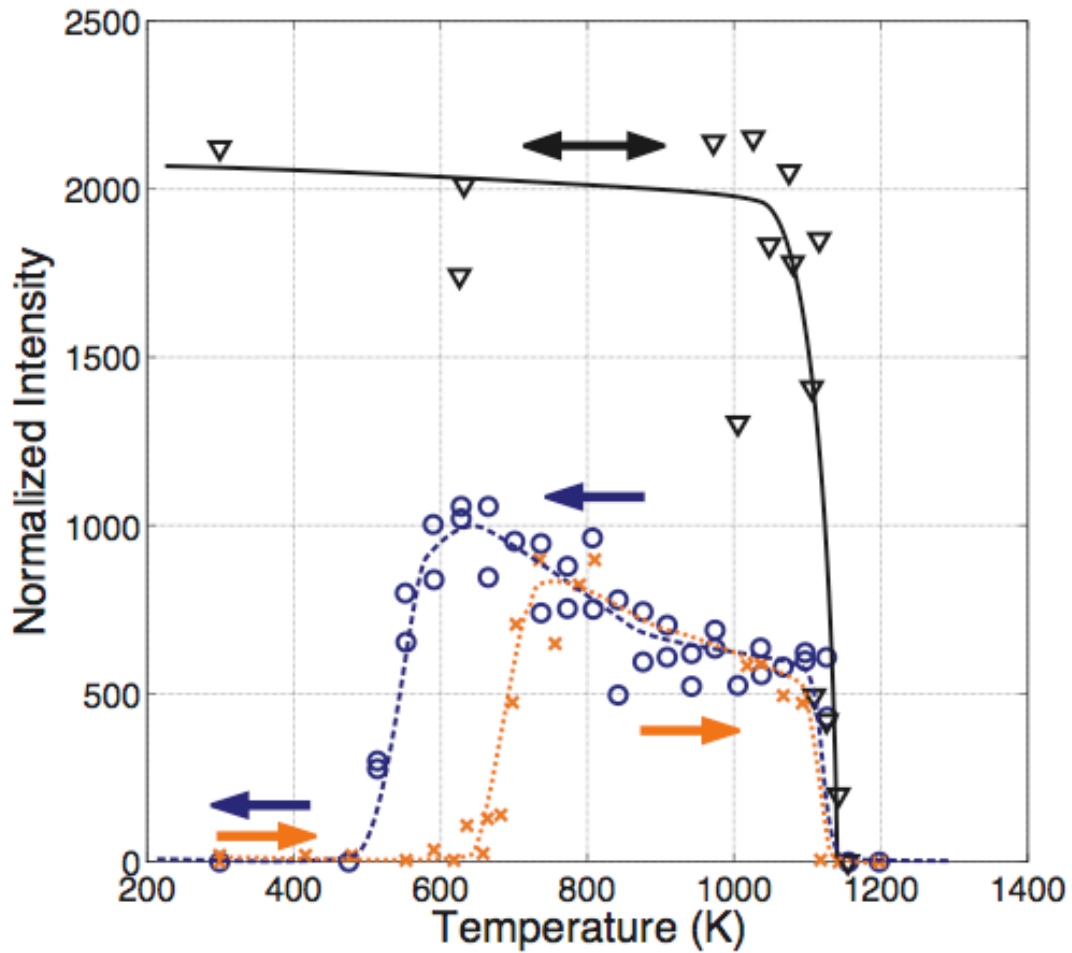

Figure 2. The hysteresis of the hex peak intensity. The (black) curve with the triangles shows the intensity in the absence of CO gas. The (blue) curve with open circles shows the intensity in the presence of 2.5% CO gas with decreasing temperature. The (orange) curve with the × marks shows the system also in the presence of CO, but upon raising the temperature from 300 K.

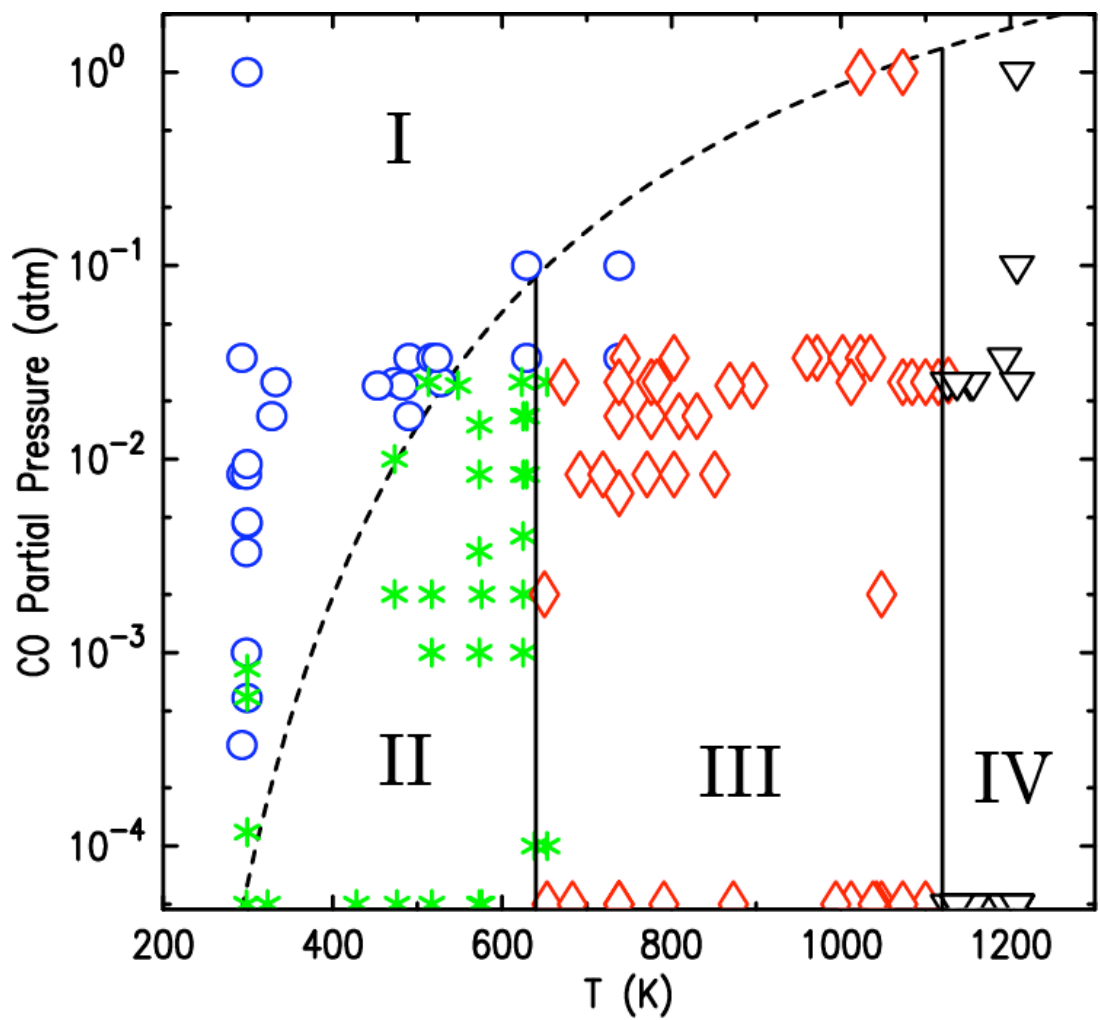

Figure 3. Phase diagram for the Au (001) surface as a function of CO partial pressure and temperature. I) CO induced (1×1), II) irreversible hex where the amount of hex does not recover after CO is removed, III) reversible hex where the hex phase recovers when the CO partial pressure is reduced, and IV) high temperature (1×1).

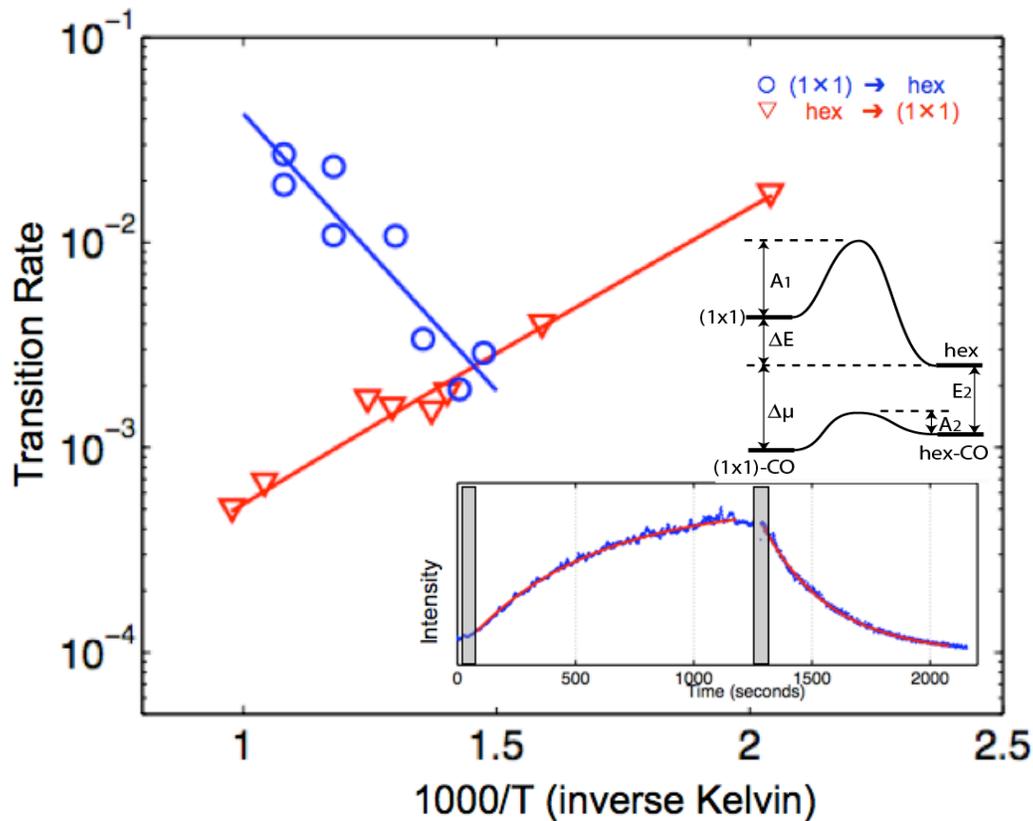

Figure 4. The transition rates for growth and decay of the hex phase. The hex to (1×1) data shown is for 3.3% CO partial pressure. The (1×1) to hex transition is in the absence of CO. The upper inset shows the energy level diagram similar to one in Ref. [11]. The lower inset shows an example of the transient data and fits from which the transition rates are obtained. The gray regions indicate the time over which the sample temperature is changing.

[1] M. Haruta, *Catal. Today* **36**, 153 (1997).

[2] M. Valden, X. Lai, D. W. Goodman, Science **281**, 1647 (1998).

[3] R.J. Behm, P.A. Thiel, P.R. Norton, and G. Ertl. J. Chem. Phys. **78**, 7437, (1983); P.A. Thiel, R.J. Behm, P.R. Norton, and G. Ertl. J. Chem. Phys. **78**, 7448, (1983).

[4] M.A.Van Hove, R. J. Koestner, P. C. Stair, J. P. Biberian, L. L. Kesmodel, I. Bartos, G. A. Somorjai, Surf. Sci., 103, 189 (1981); F. Ercolessi, E. Tossatti, and M. Parrinello. Phys. Rev. Lett. **57**, 719 (1986).

[5] S. G. J. Mochrie, D. M. Zehner, B. M. Ocko, and D. Gibbs. Phys. Rev. Lett. **64**, 2925 (1990); D. Gibbs, B. M. Ocko, D. M. Zehner, and S. G. J. Mochrie. Phys. Rev. B **42**, 7330 (1990).

[6] H. You J. Appl. Cryst. 32 614 (1999).

[7] A. Menzel, K.-C. Chang, V. Komanicky, Y. V. Tolmachev, A. V. Tkachuk, Y. S. Chu, and H. You, Phys. Rev. B 75, 035426 (2007).

[8] F.C. Nix and D. MacNair. Phys. Rev. **60**, 597, (1941).

[9] K.F. Peters, P. Steadman, H. Isern, J. Alvarez, S. Ferrer, Surf. Sci. 467, 10 (2000).

[10] K.A. Jackson, "*Kinetic Processes.*" WILEY-VCH (2004).

[11] Y. Y. Yeo; Wartnaby, C. E.; King, D. A. Science, **268**, 1731 (1995).

[12] J. Oudar, "*Physics and Chemistry of Surfaces*" Blackie, London, (1975)

[13] P. van Beurden and G. J. Kramer, J. Chem. Phys. **121**, 2317 (2004); P. van Beurden, B.S. Bunnik, G.J. Kramer, and A. Borg, Phys. Rev. Lett. **90**, 066106 (2003).

[14] A. Hopkinson, J.M. Bradley, X-C. Guo, and D.A. King. Phys. Rev. Lett. **71**, 1597 (1993); A. Hopkinson, X-C. Guo, J.M. Bradley, and D.A. King, J. Chem. Phys. **99**, 8262 (1993).

[15] V.N. Kuzovkov, O. Kortlüke, and W. von Niessen. Phys. Rev. Lett. **83**, 1636 (1999); V.N. Kuzovkov, O. Kortlüke, and W. von Niessen. Phys. Rev. E **66**, 011603 (2002).

[16] E.D.L. Rienks, G.P. van Berkel, J.W. Bakkes, R.E. Nieuwenhuys. Surf. Sci. 571, 187 (2004)

[17] P. Pyykkö, Angew. Chem. Int. Ed, 43, 4412 (2004)

[18] W.L. Yim, T. Nowitzki, M. Necke, H. Schnars, P. Nickut, J. Biener, M.M. Biener, V. Zielasek, K. Al-Shamery, T. Kluner, and M. Baumer. J. Phys. Chem. C **111**, 445 (2007).

[19] Y. Jugnet, F. J. Cadete Santos Aires, C. Deranlot, L. Piccolo, and J. C. Bertolini, Surf. Sci. 521, L639 (2002); L. Piccolo, D. Loffreda, F. J. Cadete Santos Aires, C. Deranlot, Y. Jugnet, P. Sautet and J. C. Bertolini, Surf. Sci. 566-568, 995 (2004).